\documentclass[signlecolumn,preprint]{aastex63}

\usepackage{soul}
\setstcolor{red}

\received{Xxxx xx, 2020}
\revised{Xxxx xx, 2020}
\accepted{Xxxx xx, 2020}
\submitjournal{ApJL}

\shorttitle{N$_2$ Pressure and Habitable Zone}
\shortauthors{Zhang and Yang}
\graphicspath{{./}{figures/}}

\begin{document}



\title{How does Background Air Pressure Influence the Inner Edge of the Habitable Zone for Tidally Locked Planets in a 3D View?}



\correspondingauthor{Jun Yang}
\email{junyang@pku.edu.cn}

\author[0000-0001-6194-760X]{Yixiao Zhang}
\affiliation{Dept. of Atmospheric and Oceanic Sciences, School of Physics, Peking University, Beijing 100871, China}

\author[0000-0001-6031-2485]{Jun Yang}
\affiliation{Dept. of Atmospheric and Oceanic Sciences, School of Physics, Peking University, Beijing 100871, China}

\begin{abstract}

We examine the effect of varying background N$_2$ surface pressure (labelled as $p$N$_2$) on the inner edge of the habitable zone for 1:1 tidally locked planets around M dwarfs, using the three-dimensional (3D) atmospheric general circulation model (AGCM) ExoCAM.
In our experiments, the rotation period is fixed when varying the stellar flux, in order to more clearly isolate the role of $p$N$_2$.
We find that the stellar flux threshold for the runaway greenhouse is a non-monotonous function of $p$N$_2$. This is due to the competing effects of five processes: pressure broadening, heat capacity, lapse rate, relative humidity, and clouds. These competing processes increase the complexity in predicting the location of the inner edge of the habitable zone. For a slow rotation orbit of 60 Earth days, the critical stellar flux for the runaway greenhouse onset is 1700--1750, 1900--1950, and 1750--1800~W\,m$^{-2}$ under 0.25, 1.0, and 4.0 bar of $p$N$_2$, respectively, suggesting that the magnitude of the effect of $p$N$_2$ is within $\approx$13\%. For a rapid rotation orbit, the effect of varying $p$N$_2$ on the inner edge is smaller, within a range of $\approx$7\%. Moreover, we show that Rayleigh scattering effect as varying $p$N$_2$ is unimportant for the inner edge due to the masking effect of cloud scattering and to the strong shortwave absorption by water vapor under hot climates. Future work using AGCMs having different cloud and convection schemes and cloud-resolving models having explicit cloud and convection are required to revise this problem.

\end{abstract}



\section{Introduction} \label{sec_introduction}
Various factors can influence the width of the habitable zone around stars, including stellar spectrum, planetary rotation, radius and gravity, orbital obliquity and eccentricity, air mass and composition, surface land and sea configurations, etc. \citep{Kasting_1993,Kasting_2014,Selsis_2007,Pierrehumbert_2010,Zsom_2013,Leconte_2013,Yang_2013,Yang_2014,Wolf_2014,Wolf_2015,Wordsworth_2015,Wang_2016,Shields_2016,Kopparapu_2016,Kopparapu_2017,Wolf_2017_constrains,Bin_2018,Way_2018,Yang_2019_effects,Ramirez_2020}. In this study, we investigate the effect of varying $p$N$_2$. N$_2$ is a common 
atmospheric composition of rocky planets in the solar system. The value of $p$N$_2$ is 0.78 bar on modern Earth, may less than 0.5 bar on early Earth \citep{Marty_2013,Som_2016}, 1.4 bar on Titan ($\approx$10 times Earth's value  for per unit area, given Titan's gravity is only 1.35 m\,s$^{-2}$), and 3.3 bar on Venus \citep{Ingersoll_2013}. Planets beyond the solar system are expected to also have a wide range of $p$N$_2$, which is determined by many processes such as accretion from the protoplanetary disk, impacts, lightning, volcanism, atmospheric escape, photochemistry, and ocean chemistry \citep{Johnson_2015,Wordsworth_2016,Hu_2019}.

\begin{figure}
    \includegraphics[width=0.95\textwidth]{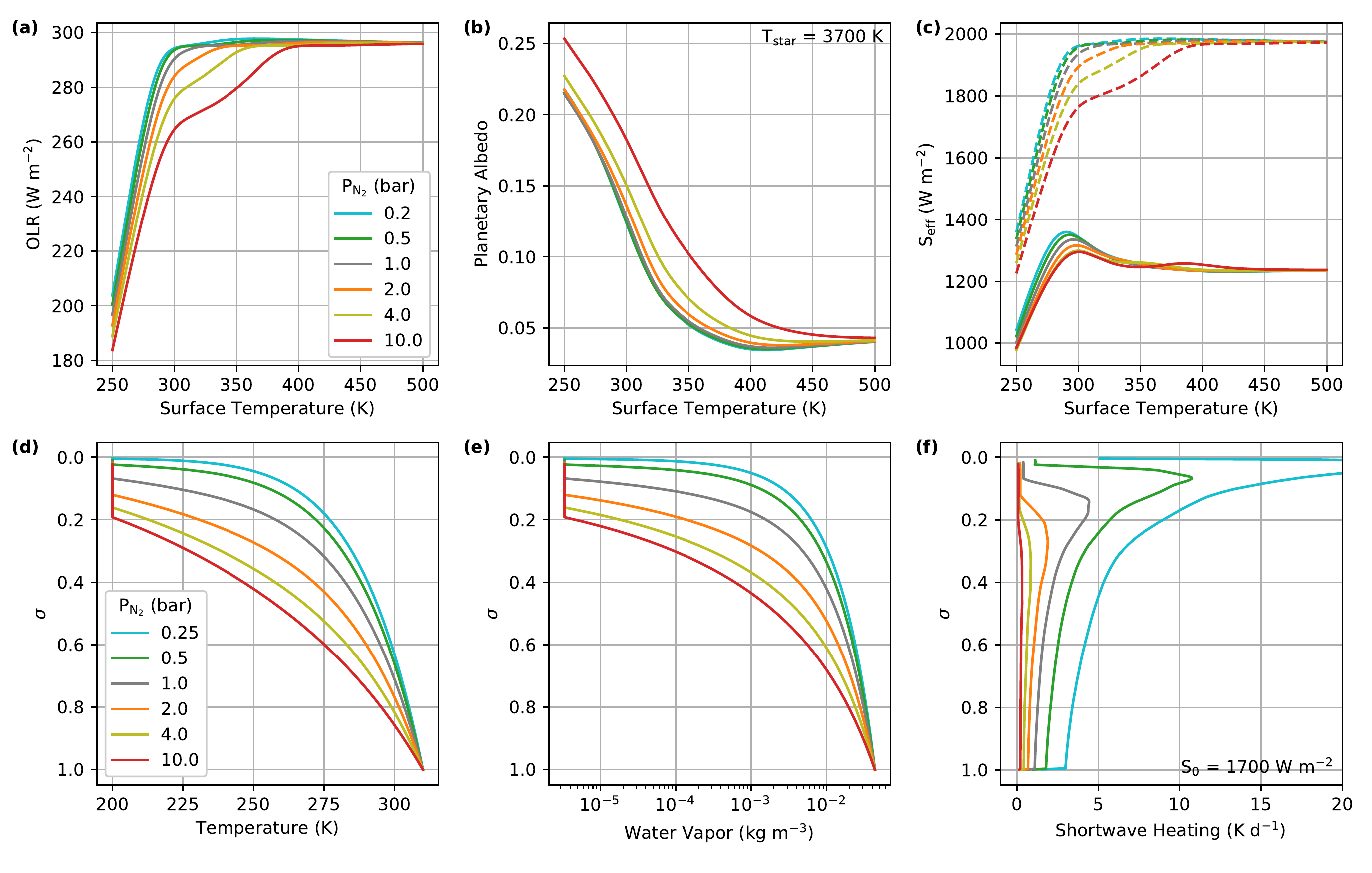}
    \caption{Effects of different $p$N$_2$ on radiation transfer, calculated using the 1D radiative-convective model ExoRT. (a--c): The effects of $p$N$_2$ on outgoing longwave radiation
        (a, $OLR$), planetary albedo (b, $\alpha_p$), and the effective
        stellar flux (defined as $(4 \times OLR)/(1-\alpha_p)$, which is the
        stellar flux required to maintain a given surface temperature; the
        dashed lines show the results under an assumed $\alpha_p$ of 0.4),
        as a function of surface temperature from 250 to 500~K. (d--f):
        Vertical profiles of air temperature in $\sigma$-coordinate under
        a fixed surface temperature of 310~K (d), water vapor density (e),
        and calculated shortwave heating rate (f). In these calculations,
        stellar flux is 1700 W\,m$^{-2}$, star temperature is 3700 K, solar
        zenith angle is 60$^{\circ}$, the surface albedo is 0.25, and no cloud
        is included. The air temperature decreases from the surface to the
        top following moist adiabatic until it reaches 200~K, above which
        the atmosphere is isothermal with a constant water mixing ratio
        (see~\cite{Kasting_1993}). The decreasing of shortwave
        heating rate with $p$N$_2$ in (f) is due to the combined effect of
        the reduction of water vapor aloft and the increase of heat capacity; the effect of heat capacity dominates.
    }
    \label{fig_air_mass_effects}
\end{figure}

Although N$_2$ is not a greenhouse gas, it can influence planetary climate through several processes, including pressure broadening (as well as collision-induced N$_2$--N$_2$ continuum absorption; warming effect; Fig.~\ref{fig_air_mass_effects}(a)), Rayleigh scattering (cooling effect, Fig.~\ref{fig_air_mass_effects}(b)), heat capacity, lapse rate (i.e., the vertical profile of air temperature), and energy transport. The relative importance of these effects depends on the level of $p$N$_2$ and the climate state. For temperate climates of early Earth and early Mars for which $p$N$_2$ is not very high, the warming effect of pressure broadening dominates \citep{Goldblatt_2009,Von_2013_n2,Charnay_2013,Wolf_2014}. For temperate or cold climate with high-level $p$N$_2$ but relatively low greenhouse gases (such as H$_2$O and CO$_2$), the cooling effect of Rayleigh scattering dominates \citep{Keles_2018}. Moist adiabatic lapse rate ($-\partial{}T/\partial{}z$) increases with $p$N$_2$ (Fig.~\ref{fig_air_mass_effects}(d)), which influences vapor concentration aloft (Fig.~\ref{fig_air_mass_effects}(e)), the strength of greenhouse effect \citep{Nakajima_1992, Pierrehumbert_2010}, and shortwave heating rate (Fig.~\ref{fig_air_mass_effects}(f)). Atmospheric heat capacity ($C_p$d$m$, where $C_p$ is the specific heat capacity and d$m$ is the air mass per unit area between two adjacent vertical levels) increases with $p$N$_2$, which can also strongly affect shortwave heating rate, longwave cooling rate, and condensation heating rate \citep{Chemke_2016}. As shown in Fig.~\ref{fig_air_mass_effects}(f), the shortwave heating rate (=$F_{SW}$/($C_p$d$m$), where $F_{SW}$ is the net shortwave flux for each level) decreases significantly with $p$N$_2$ due to the decrease of water vapor aloft and the increase of heat capacity. The magnitude of $p$N$_2$ can also influence horizontal and vertical energy transports \citep{Kaspi_2015,Chemke_2016,Chemke_2017,Komacek_2019}. In this work, we examine the net effect of $p$N$_2$ on the inner edge of the habitable zone with a model including all these processes as well as clouds and atmospheric sub-saturation.

Following \cite{Kasting_1993}, the inner edge is defined as the location where absorbed shortwave radiation ($ASR$) of the planet exceeds the upper limit of outgoing longwave radiation at the top of the atmosphere ($OLR_m$) with the entire ocean would evaporate. The onset of a moist greenhouse state (i.e., high water vapor concentration above the tropopause and significant water loss to space, see \cite{Wordsworth_2014}) will not be considered in this study, because the work of \cite{Kopparapu_2017} and \cite{Fujii_2017} showed that slow-rotation tidally locked planets around low mass stars can undergo water loss but remain habitable. Moreover, for water-rich planets, only a fraction of the ocean may be lost within the lifetime of the planets, so water loss in moist greenhouse state is not a ultimate threat for planetary habitability \citep{Selsis_2007,Moore_2020}. Here, we focus on the runaway greenhouse.

Using 1D radiative-convective model, \cite{Kasting_1993} showed that varying $p$N$_2$ has an insignificant effect on the runaway greenhouse limit (see their Table~2). This is due to that for the runaway greenhouse state the atmosphere is dominated by water vapor and the presence of N$_2$ is not so important. Studies with updated absorption coefficients for CO$_2$ and H$_2$O, found that varying $p$N$_2$ has an effect of within $\approx$10\% on the runaway greenhouse limit (Fig.~4.38 in \cite{Pierrehumbert_2010}; Fig.~5 in \cite{Goldblatt_2013}, Fig.~1(a) in \cite{Kopparapu_2014}, and see also Fig.~\ref{fig_air_mass_effects}(c) here), due to the combined effects of pressure broadening, lapse rate, and Rayleigh scattering. \cite{Ramirez_2020} further showed that the effect of $p$N$_2$ for M and K dwarfs is much weaker than that for F and A stars, due to the lower Rayleigh scattering and higher near-infrared absorption of water vapor under redder spectra. \cite{Vladilo_2013} and \cite{Zsom_2013} found that the effect of $p$N$_2$ on the inner edge could be as large as 65\% in 1D climate calculations, however, they used boiling point as the inner edge (this is unconventional in literature) and meanwhile they employed a relative humidity of 1\% or 50\% rather than 100\% that is always assumed in 1D radiative-convective models.

Two weaknesses of the 1D models are that clouds are not simulated and relative humidity is fixed, because clouds and humidity are primarily determined by 3D atmospheric circulation and convection. In this study, we plan to improve the understanding of this problem through 3D simulations with an AGCM (Section~\ref{sec_methods}). We focus on tidally locked planets around M dwarfs due to their relatively large planet-to-star ratio and frequent transits. Previous 3D experiments have been employed to examine the inner edge for tidally locked planets (such as \cite{Yang_2013,Yang_2014,Yang_2019,Yang_2019_simulations,Way_2016,Kopparapu_2016,Kopparapu_2017,Bin_2018}), but these studies always assumed $p$N$_2$ being equal to $\approx$1.0~bar. Below, we show that the magnitude of varying $p$N$_2$ on the runaway greenhouse limit is within $\approx$13\%, similar to that found in 1D radiative-convective models, but the underlying mechanisms are different and more complex than that found in 1D models. More important, we find that the dependence of the inner edge on $p$N$_2$ is non-monotonic, especially for a slow rotation orbit (Section~\ref{sec_results}). A summary is shown in Section~\ref{summary}.



\section{Methods}\label{sec_methods}

The atmospheric general circulation model used in this study is ExoCAM. The model is based on the Community Atmospheric Model version 4 but modified by Eric Wolf for simulating early Earth and terrestrial exoplanets \citep{Wolf_2014,Wolf_2015,Wolf_2017_constrains,Wolf_2017_assessing}. Two main modifications are the radiation transfer for high concentrations of CO$_2$ and H$_2$O and the numerical solver for entropy calculation within the Zhang-MacFarlane convection parameterization, so that ExoCAM is able to well simulate the onset of runaway greenhouse. In the model, correlated-$k$ coefficients are based on the database of HITRAN 2012, and water-vapor continuum absorption and collision-induced N$_2$-N$_2$ absorption are included.

Atmospheric compositions are set to Earth-like, including N$_{2}$ and H$_{2}$O, but O$_2$, O$_3$, CO$_{2}$, CH$_{4}$ and aerosols are not considered. The total pressure of the atmosphere is given by $P_\mathrm{tot} = p\mathrm{N}_2 + p\mathrm{H}_\mathrm{2}\mathrm{O}$. Several different surface $p$N$_2$ were examined, from 0.25 to 10 bar (Table~\ref{tab_all_runs}). Planetary radius and gravity are the same as Earth. Two different rotation periods are tested, 60 and 6 Earth days. All the experiments are set to be 1:1 tidally locked, i.e., rotation period\,$=$\,orbital period. The stellar temperature is 3700~K for the slow rotation orbit and 2600~K for the rapid rotation orbit. Stellar spectra are from the BT\_Settl stellar model \citep{Allard_2007}. A series of stellar flux were examined with an interval of 50 or 100~W\,m$^{-2}$ (Table~\ref{tab_all_runs}). Our experimental design is different from that employed in \cite{Kopparapu_2016,Kopparapu_2017}, who modified the rotation period and the stellar flux simultaneously. Their experiments are able to self-consistently consider the combined effect of the Coriolis force and stellar flux, but do not allow separate considerations of each factor. In our simulations, the rotation period is fixed when varying the stellar flux, as was done by such as \cite{Merlis_2010}, \cite{Way_2016}, \cite{Noda_2017}, and \cite{Bin_2018}. This design allows us to isolate the effect of stellar radiation and it is sufficient to demonstrate the role of varying air pressure. Future work is required to consistently investigate the combined effect of varying the rotation rate and stellar flux.

\begin{table}
    \centering
    \caption{Global-mean surface temperature ($T_s$), energy balance at the top of the atmosphere (net shortwave minus net longwave, labelled as EB), planetary albedo ($\alpha{}_p$), surface albedo ($\alpha{}_s$), vertically integrated water vapor ($Q$), clear-sky greenhouse effect ($G_c$), longwave cloud radiation effect (LWCRE), and shortwave cloud radiation effect (SWCRE) under different stellar temperatures ($T_{\mathrm{star}}$), rotation periods ($P$), surface N$_2$ pressures ($p$N$_2$), and stellar fluxes ($S_0$), simulated with ExoCAM. Averages of the last 10 model (earth) years in each experiment are showed, except in the runaway greenhouse states (red color) averages of the last one model year before model crash are listed.}
    \label{tab_all_runs}
    \footnotesize{

\begin{tabular}{cccccccccccc}
    \hline
    \hline
    $T_\mathrm{star}$ &$P$    &$p$N$_2$                     &$S_0$                 &$T_s$                            &EB                               &$\alpha_p$            &$\alpha_s$            &$Q$                                 &$G_\mathrm{c}$                   &LWCRE                  &SWCRE\\
    (K)               &(days) &(bar)                     &(W\,m$^{-2}$)         &(K)                              &(W\,m$^{-2}$)                    &(0-1)                 &(0-1)                 &(kg\,m$^{-2}$)                     &(K)                              &(W\,m$^{-2}$)         &(W\,m$^{-2}$) \\
    \hline
    3700              &60     &\phantom{0}0.25           &1600                  &\phantom{$>$}269.3               &\phantom{00}\llap{-}2.0          &0.39                  &0.08                  &\phantom{000}42.6                  &\phantom{00}7.8                  &21.4                  &-131.4 \\
                      &       &                          &1700                  &\phantom{$>$}277.6               &\phantom{00}\llap{-}0.3          &0.40                  &0.06                  &\phantom{000}63.7                  &\phantom{0}12.0                  &28.5                  &-150.8  \\
                      &       &                          &\textcolor{red}{1750} &\textcolor{red}{326.0 (runaway)} &\textcolor{red}{\phantom{0}70.9} &\textcolor{red}{0.19} &\textcolor{red}{0.05} &\phantom{0}\textcolor{red}{1536.0} &\phantom{0}\textcolor{red}{50.5} &\textcolor{red}{34.4} &\phantom{0}\textcolor{red}{-72.2} \\
    \cline{3-12}
                      &       &\phantom{0}0.5\phantom{0} &1700                  &\phantom{$>$}265.8               &\phantom{00}\llap{-}1.8          &0.43                  &0.11                  &\phantom{00}37.6                   &\phantom{00}7.1                  &17.2                  &-159.1 \\
    \cline{3-12}
                      &       &\phantom{0}1.0\phantom{0} &1600                  &\phantom{$>$}258.9               &\phantom{00}\llap{-}1.5          &0.44                  &0.16                  &\phantom{000}26.7                  &\phantom{00}5.1                  &16.4                  &-147.3 \\
                      &       &                          &1700                  &\phantom{$>$}264.8               &\phantom{00}\llap{-}1.0          &0.44                  &0.13                  &\phantom{000}37.2                  &\phantom{00}7.4                  &15.3                  &-161.4 \\
                      &       &                          &1750                  &\phantom{$>$}270.3               &\phantom{00}\llap{-}0.5          &0.44                  &0.09                  &\phantom{000}49.2                  &\phantom{0}10.0                  &19.5                  &-170.5 \\
                      &       &                          &1800                  &\phantom{$>$}278.9               &\phantom{00}0.4                  &0.44                  &0.07                  &\phantom{000}74.3                  &\phantom{0}14.9                  &25.6                  &-176.3 \\
                      &       &                          &1900                  &\phantom{$>$}312.1               &\phantom{00}0.8                  &0.44                  &0.07                  &\phantom{00}325.8                  &\phantom{0}41.8                  &34.8                  &-189.3 \\
                      &       &                          &\textcolor{red}{1950} &\textcolor{red}{383.6 (runaway)} &\textcolor{red}{114.0}           &\textcolor{red}{0.05} &\textcolor{red}{0.05} &\textcolor{red}{10565.5}           &\phantom{0}\textcolor{red}{93.8} &\textcolor{red}{42.1} &\phantom{0}\textcolor{red}{-11.7} \\
    \cline{3-12}
                      &       &\phantom{0}2.0\phantom{0} &1700                  &\phantom{$>$}278.2               &\phantom{00}\llap{-}0.6          &0.41                  &0.07                  &\phantom{000}67.2                  &\phantom{0}15.9                  &22.3                  &-153.1 \\
    \cline{3-12}
                      &       &\phantom{0}4.0\phantom{0} &1600                  &\phantom{$>$}282.7               &\phantom{00}\llap{-}0.1          &0.38                  &0.07                  &\phantom{000}72.5                  &\phantom{0}20.6                  &22.4                  &-124.4 \\
                      &       &                          &1700                  &\phantom{$>$}296.7               &\phantom{00}0.4                  &0.38                  &0.07                  &\phantom{00}137.4                  &\phantom{0}31.0                  &22.9                  &-137.0 \\
                      &       &                          &1750                  &\phantom{$>$}308.7               &\phantom{00}\llap{-}0.9          &0.40                  &0.07                  &\phantom{00}218.4                  &\phantom{0}41.6                  &24.1                  &-147.6 \\
                      &       &                          &\textcolor{red}{1800} &\textcolor{red}{371.5 (runaway)} &\phantom{00}\textcolor{red}{9.1} &\textcolor{red}{0.37} &\textcolor{red}{0.06} &\phantom{0}\textcolor{red}{7480.4} &\textcolor{red}{107.6}           &\textcolor{red}{39.8} &\textcolor{red}{-149.4} \\
    \cline{3-12}
                      &       &10.0\phantom{0}           &1700                  &\phantom{$>$}343.9               &\phantom{00}\llap{-}0.1          &0.38                  &0.06                  &\phantom{00}752.1                  &\phantom{0}77.1                  &22.3                  &-126.9 \\
    \hline
    2600              &6      &\phantom{0}0.25           &1200                  &\phantom{$>$}256.1               &\phantom{00}\llap{-}3.7          &0.24                  &0.14                  &\phantom{000}{32.2}                &\phantom{00}5.2                  &11.7                  &\phantom{0}-44.6\\
                      &       &                          &1300                  &\phantom{$>$}274.8               &\phantom{00}\llap{-}1.1          &0.19                  &0.07                  &\phantom{000}{91.5}                &\phantom{0}11.2                  &22.7                  &\phantom{0}-48.5\\
                      &       &                          &\textcolor{red}{1350} &\textcolor{red}{336.4 (runaway)} &\phantom{0}\textcolor{red}{19.1} &\textcolor{red}{0.09} &\textcolor{red}{0.04} &\phantom{0}\textcolor{red}{1589.6} &\phantom{0}\textcolor{red}{59.4} &\textcolor{red}{34.2} &\phantom{0}\textcolor{red}{-25.3}\\
    \cline{3-12}
                      &       &\phantom{0}1.0\phantom{0} &1200                  &\phantom{$>$}259.3               &\phantom{00}\llap{-}1.8          &0.23                  &0.14                  &\phantom{000}{25.6}                &\phantom{00}6.4                  &12.7                  &\phantom{0}-49.7\\
                      &       &                          &1300                  &\phantom{$>$}277.8               &\phantom{00}\llap{-}0.9          &0.17                  &0.07                  &\phantom{000}{69.4}                &\phantom{0}13.5                  &17.2                  &\phantom{0}-45.8\\
                      &       &                          &1350                  &\phantom{$>$}289.6               &\phantom{00}0.1                  &0.14                  &0.06                  &\phantom{00}{157.3}                &\phantom{0}19.9                  &19.4                  &\phantom{0}-38.6\\
                      &       &                          &\textcolor{red}{1400} &\textcolor{red}{363.7 (runaway)} &\phantom{0}\textcolor{red}{24.6} &\textcolor{red}{0.06} &\textcolor{red}{0.05} &\phantom{0}\textcolor{red}{4885.7} &\phantom{0}\textcolor{red}{85.1} &\textcolor{red}{85.1} &\textcolor{red}{-17.3}\\
    \cline{3-12}
                      &       &\phantom{0}4.0\phantom{0} &1200                  &\phantom{$>$}262.6               &\phantom{00}\llap{-}2.1          &0.21                  &0.16                  &\phantom{000}{16.3}                &\phantom{00}9.5                  &\phantom{0}6.1        &\phantom{0}-43.4\\
                      &       &                          &1400                  &\phantom{$>$}310.6               &\phantom{00}0.2                  &0.12                  &0.05                  &\phantom{00}{175.6}                &\phantom{0}37.4                  &13.0                  &\phantom{0}-32.7\\
                      &       &                          &\textcolor{red}{1450} &\textcolor{red}{359.3 (runaway)} &\phantom{00}\textcolor{red}{6.6} &\textcolor{red}{0.06} &\textcolor{red}{0.06} &\phantom{0}\textcolor{red}{2599.9} &\phantom{0}\textcolor{red}{79.8} &\textcolor{red}{14.5} &\phantom{0}\textcolor{red}{-19.4} \\
    \hline
\end{tabular}

}
\end{table}

The atmosphere is coupled to an immobile, slab ocean with a depth of 50 m and with no oceanic dynamics; no continent is considered. Previous studies have shown that oceanic heat transport is important for the planets in the middle range of the habitable zone, but its magnitude is much smaller for planets close to the inner edge \citep{Yang_2019_ocean}. The latter is due to that under hot climates, temperature contrasts between dayside and nightside 
are small and thereby surface winds are weak. Sea ice is allowed to form when the surface temperature is below the freezing point (271.35~K), and the albedos of sea ice and snow depend on the stellar spectrum \citep{Wolf_2014}. Horizontal resolution of the model is 4$^{\circ}$ in latitude and 5$^{\circ}$ in longitude with 40 levels in the vertical direction, and the top of the model is $\approx$1~hPa. The time step is 30 minutes. Each experiment was run for tens of to one hundred Earth years until the surface and atmosphere reach equilibrium. By default, averages of the last ten years were used here.



\section{Results} \label{sec_results}

\subsection{Non-monotonic Dependence of Planetary Climate on \texorpdfstring{pN$_2$}{pN2}}\label{sec_climate}

\begin{figure}
    \includegraphics[width=1.08\linewidth]{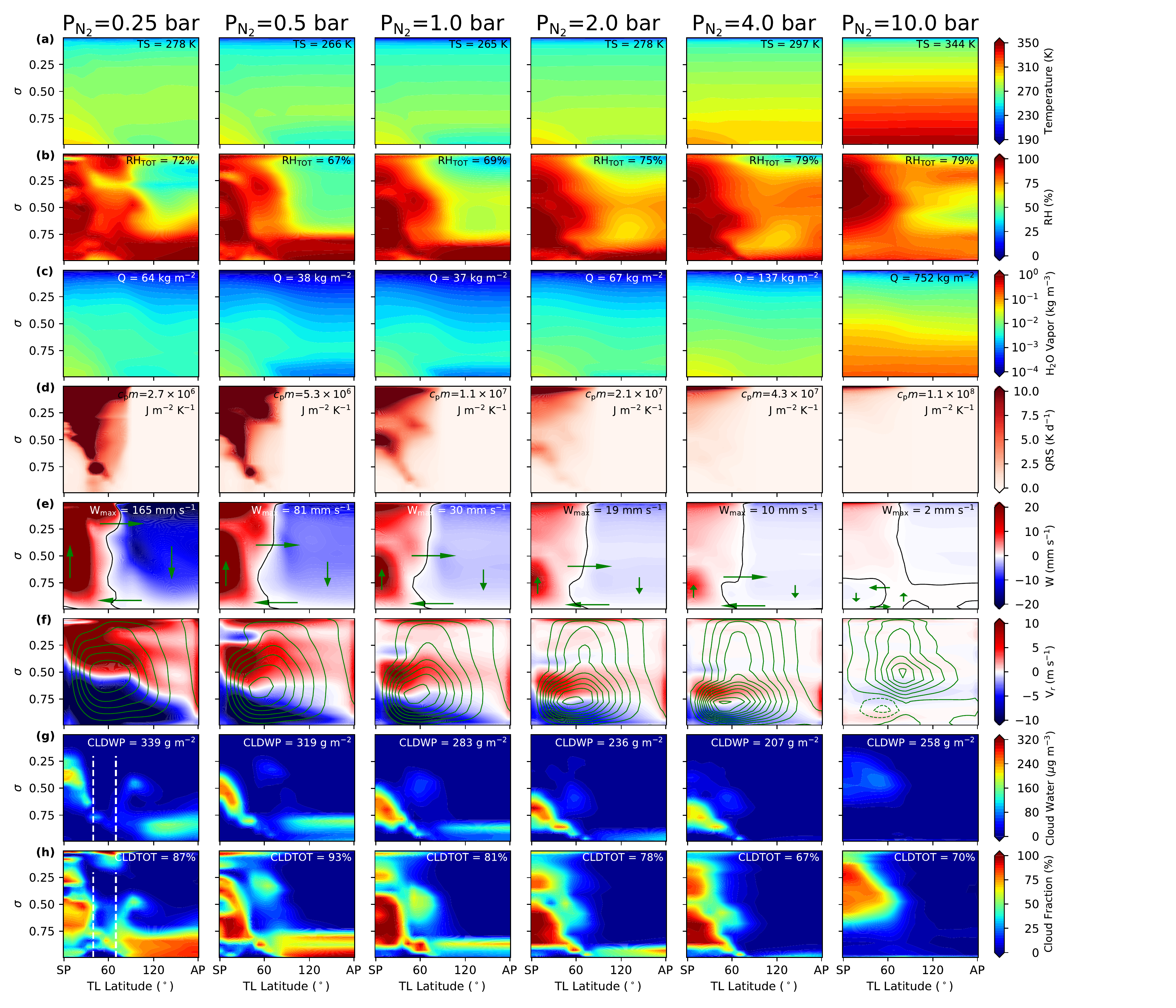}
    \caption{Effects of $p$N$_2$ on the climate of a tidally locked aqua-planet. (a) air temperature, (b) relative humidity (RH), (c) water vapor concentration, (d) shortwave heating rate (QRS), (e) vertical velocity (W, solid line is zero velocity), (f) radial velocity (V$_r$), (g) cloud water content, and (h) cloud fraction in tidally-locked (TL) coordinates, for $p$N$_2$ of 0.25, 0.5, 1.0, 2.0, 4.0, and 10.0 bar from left to right columns. The substellar point (SP) and anti-stellar point (AP) are at $0^\circ{}$ and $180^{\circ}$, respectively. The contour lines in (f) are mass streamfunction with intervals of 10$^{11}$ kg\,s$^{-1}$ (solid lines: clockwise; dashed lines: anti-clockwise). The vertical dashed lines in (g-h) mask the region where the cloud fraction is relatively low. The numbers in the right corner of each panel is global-mean surface temperature in (a), total relative humidity in ((b), defined as the percentage of water vapor by mass contained in the whole atmosphere compared with the vapor mass that the atmosphere could theoretically hold if saturated everywhere, following \cite{Wolf_2015}), vertically integrated water vapor amount in (c), total atmospheric heat capacity in ((d), defined as C$_pm$ where C$_p$ is the specific heat capacity and $m$ is the vertically integrated air mass per unit area), maximum vertical velocity below $\sigma=0.1$ in (e), vertically integrated cloud water path in (g), and total cloud water fraction in ((h), assuming maximum--random overlap). The stellar flux is 1700~W\,m$^{-2}$, star temperature is 3700 K, and rotation period is 60 Earth days in all these experiments.}
    \label{fig_air_mass_climate}
\end{figure}

The planetary climate is a non-monotonous function of $p$N$_2$, under a given stellar flux. As shown in Fig.~\ref{fig_air_mass_climate} and Table~\ref{tab_all_runs}, the global-mean surface temperature is 278, 266, 265, 278, 297, and 344~K for $p$N$_2$ of 0.25, 0.5, 1.0, 2.0, 4.0, and 10.0 bar, respectively. Several competing processes cause the non-monotonicity, including pressure broadening, lapse rate, relative humidity, heat capacity, and clouds.

For a higher $p$N$_2$ than 1.0 bar, such as 4.0 bar, the warmer surface mainly results from three responses: stronger pressure broadening (and N$_2$--N$_2$ absorption), higher relative humidity, and a lower planetary albedo. As $p$N$_2$ is increased, the global-scale `Walker circulation' (with upwelling over the substellar region and downwelling in the rest region) becomes shallower in altitude, weaker in strength (W and V$_r$ winds), and meanwhile the upwelling area becomes wider in horizontal scale and the downwelling area becomes narrower (Fig.~\ref{fig_air_mass_climate}(e)). The reduction of the downwelling area and the weakening of the downwelling strength allow the atmosphere to be more saturated, so the relative humidity increases especially in the upper troposphere (Fig.~\ref{fig_air_mass_climate}(b)). The total relative humidity is 69\% and 79\% in the cases of 1.0 and 4.0 bar, respectively. So, the atmosphere can hold more water vapor and have a stronger greenhouse effect, warming the surface. Water vapor feedback further increases clear-sky atmospheric greenhouse effect and amplifies the surface warming. The weakening of atmospheric circulation with $p$N$_2$ is consistent with the ideal 3D simulations of \cite{Kaspi_2015, Chemke_2016} and \cite{Chemke_2017}. Note that the mass streamfunction (contour lines in Fig.~\ref{fig_air_mass_climate}(f)) becomes stronger with $p$N$_2$ due to the increase in air mass, despite of the weaker velocities. The stronger mass streamfunction transports more heat from the dayside to the nightside, reducing the day-to-night thermal contrast (Fig.~\ref{fig_air_mass_climate}(a)).

In the case of 4.0 bar, planetary albedo is 0.38 while it is 0.44 in the case of 1.0 bar (Table~\ref{tab_all_runs}). This implies that Rayleigh scattering, which increases with $p$N$_2$, is not the reason. The planetary albedo on tidally locked planets is mainly from the strong convection and clouds over the substellar region (Fig.~\ref{fig_air_mass_climate}(g \& h), \cite{Yang_2013}), which weakens the effect of Rayleigh scattering. Moreover, at high temperatures, near-infrared absorption by water vapor also reduces the effect of Rayleigh scattering \citep{Ramirez_2020}; as shown in Fig.~\ref{fig_air_mass_effects}(b), planetary albedos under different $p$N$_2$ approach to a constant value under hot climates in the 1D radiative-convective model without clouds. Therefore, we can say that Rayleigh scattering is unimportant for the inner edge of the habitable zone. The lower planetary albedo in the case of 4.0 bar is related to the reduction of cloud water amount, which exhibits a decreasing trend with $p$N$_2$ (Fig.~\ref{fig_air_mass_climate}(g) except the 10.0 bar case). The underlying mechanism is the increase of lapse rate (going closer to dry adiabatic) when $p$N$_2$ is increased. Therefore, less condensation occurs in the troposphere. The cloud fraction, determined by the combination of convection mass flux, stratification, and relative humidity \citep{Neale_2010}, does not exhibit a clear trend (Fig.~\ref{fig_air_mass_climate}(h)). Because convection and clouds are parameterized in the model, the response of clouds to varying $p$N$_2$ may be model-dependent; large differences exist in simulating clouds among AGCMs as shown in \cite{Yang_2019_simulations} and \cite{Fauchez_2020}.

The 10.0~bar case exhibits a quite different behaviour: the global-scale Walker cell and the convection occur in the levels above 7.0 bar, below which temperature gradients are small even around the terminators and the atmosphere is calm (the rightest panels in Fig.~\ref{fig_air_mass_climate}); this climate state is similar to that on Venus \citep{Read_2018}. In this case, surface temperature differences between the day and night are within 3~K, and the low-level atmosphere shows an anti-clockwise (rather than clockwise) circulation. The latter is due to the weakening of convective fluxes with increasing $p$N$_2$ \citep{Chemke_2017} and to the cooling effect of the evaporation of precipitating droplets over the substellar region below the level of 7.0 bar (figure not shown), which causes atmospheric downwelling there.

For a lower $p$N$_2$ than 1.0 bar, such as 0.25 bar, the relatively warmer surface is mainly from cloud response and lapse rate change. Planetary albedos are 0.40 and 0.44 in the cases of 0.25 and 1.0 bar, respectively. The lower albedo under 0.25 bar is mainly from the smaller cloud fraction in the region between 50$^\circ$ and 80$^\circ$ (in tidally locked coordinates, see the leftmost panels in Fig.~\ref{fig_air_mass_climate}(g-h)). This region is dominated by down-welling under $p$N$_2$\,=\,0.25~bar rather than upwelling as that in $p$N$_2$\,=\,1.0 bar, which is related to the expending trend of the upwelling region with increasing $p$N$_2$ as described above. For a lower $p$N$_2$, the lapse rate is smaller under a given surface temperature, so that the air temperature aloft is higher (Fig.~\ref{fig_air_mass_effects}(d)) and more vapor can be maintained in the atmosphere (Fig.~\ref{fig_air_mass_effects}(e)), following the Clausius–Clapeyron relation. So, the 0.25 bar case can hold more water vapor in the air, warming the surface. Again, water vapor feedback acts to further amplify the surface warming. The total relative humidity in the 0.25~bar case is also higher than that in the 1.0~bar case but with a much smaller magnitude, 72\% versus 69\%, which is likely due to the increase of vapor concentration aloft, associated with the reduced lapse rate.

A clear trend is the decreasing of shortwave heating rate with increasing $p$N$_2$ (Fig.~\ref{fig_air_mass_climate}(d)). It is due to the increases of heat capacity and lapse rate with $p$N$_2$, as described in Section~\ref{sec_introduction} and in \cite{Chemke_2016}. The effect of heat capacity dominates; for example, vertically integrated water vapor amounts are 64 and 37 kg\,m$^{-2}$ in the cases of 0.25 and 1.0 bar, respectively, while the total heat capacity of the former is only $\approx$25\% of that in the latter. Another clear trend is the reduction of low-level clouds on the nightside (Fig.~\ref{fig_air_mass_climate}(g \& h)). These clouds are trapped under the temperature inversion by large-scale circulation. As the circulation shrinks in altitude or completely collapses, these clouds become thinner or disappear. These clouds have no shortwave cooling effect due to the lack of stellar energy deposition on the nightside but have a negative longwave cloud radiation effect (LWCRE), because they are close to the temperature inversion. If oceanic heat transport is included in the simulations, the temperature inversion will be weaker or even disappear, and the value of LWCRE will turn to positive (see Fig.~\ref{fig_air_mass_climate} in \cite{Yang_2019_ocean}).

Table~1 exhibits that energy balance of the system is negative with magnitudes of 1--4 W\,m$^{-2}$, especially in the relatively cooler experiments. This energy imbalance is mainly due to continuous growths of surface snow and sea ice, especially on the nightside (figures not shown). The continued growth of sea ice is due to the fact that neither geothermal heat flux at the ocean bottom nor oceanic heat transport (OHT) from the open ocean to the ice beneath is considered in the simulations. Without OHT, the ice will grow to thousands of meters \citep{Menou_2013}. If OHT is included, the ice will be limited to several or tens of meters \citep{Hu_2014, Yang_2014_water}. However, the snow depth and ice thickness do not influence the surface temperature significantly (within 1.5~K in global mean); this is because a small thickness of snow or sea ice is able to have a strong isolation effect between the ocean and the atmosphere and to have a relatively high, saturated surface albedo in the regions under stellar deposition.


\subsection{The Inner Edge of the Habitable Zone} \label{sec_inner}

For the inner edge of the habitable zone, we focus on three levels of $p$N$_2$, 0.25, 1.0, and 4.0 bar. The stellar flux thresholds for the onset of the runaway greenhouse on a slowly rotating aqua-planet (60 earth days) are 1700--1750, 1900--1950, and 1750--1800 W\,m$^{-2}$, respectively. This again exhibits a non-monotonic feature. For a rapidly rotating planet, the stellar flux thresholds are smaller, 1300--1350, 1350--1400, and 1400--1450~W\,m$^{-2}$, respectively. The differences are smaller for the rapidly rotating aqua-planet; this is primarily due to the smaller cloud albedo (Table~\ref{tab_all_runs}).

\begin{figure}
\begin{center}
    \includegraphics[width=1.0\linewidth]{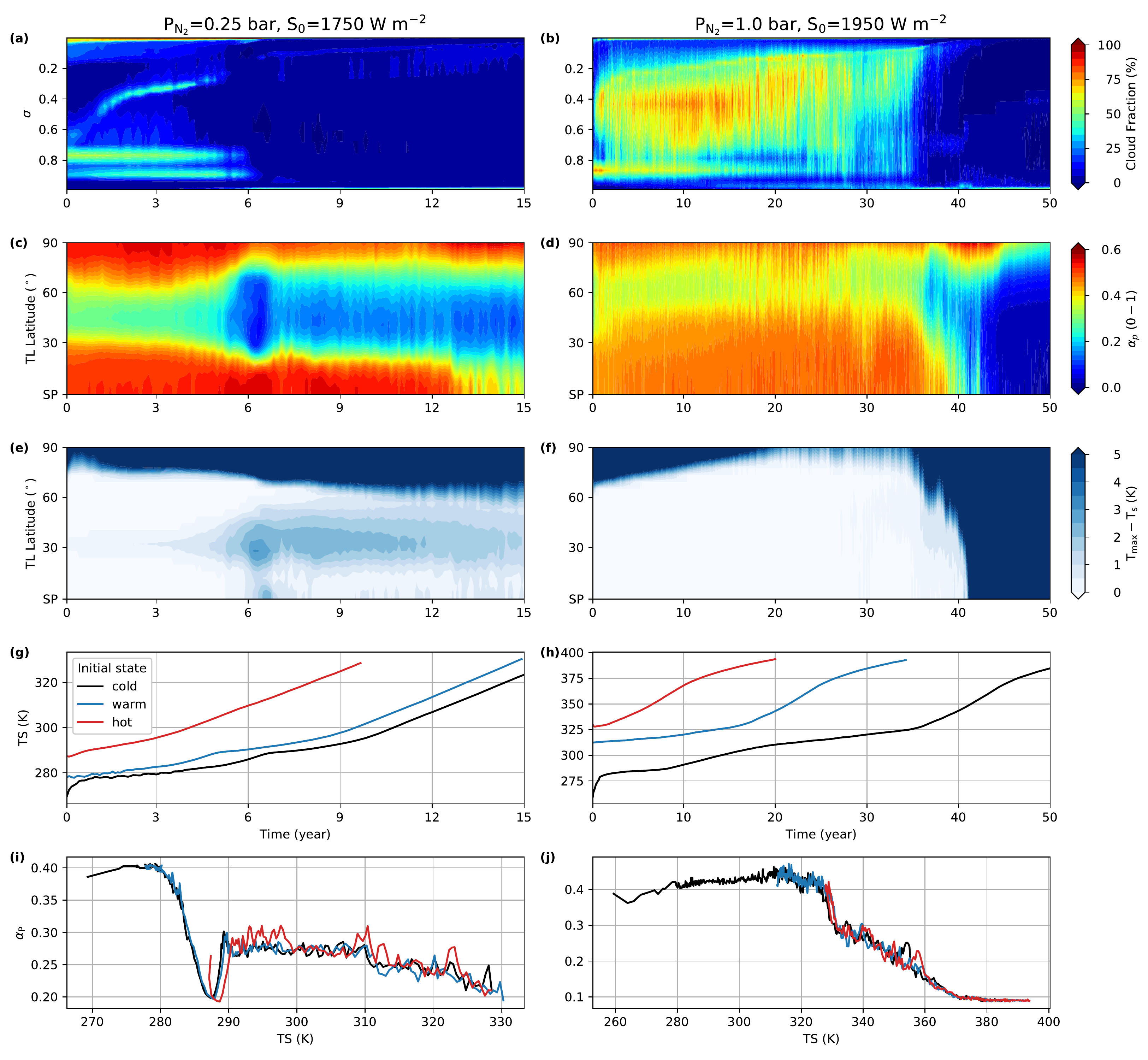}
    \caption{The trigger of runaway greenhouse (a--f) and its insensitivity to initial condition (g--j), in the slowly rotating experiments. (a-f): Time evolution of cloud fraction in $\sigma$ coordinates (a--b, averages between 0$^\circ$ and 60$^\circ$), and planetary albedo (c--d), and temperature inversion (e--f, defined as maximum air temperature minus the corresponding surface temperature) in tidally locked coordinates. SP is the substellar point. (g--j): The evolution of global-mean surface temperature as a function of time (g \& h) and the evolution of planetary albedo as a function of the global-mean surface temperature (i \& j) in the experiments using different initial states: cold, warm, and hot. Left panels are for the experiments of $p$N$_2$ of 0.25 bar and stellar flux of 1750~W\,m$^{-2}$ and right panels for 1.0 bar and 1950~W\,m$^{-2}$. The cases of 4.0 bar are similar to the 1.0 bar cases and so they are not shown for clarity. These experiments suggest that the trigger of the runaway greenhouse state is associated with the temperature inversion onset and cloud collapse, and it is insensitive to the initial condition.}
     \label{fig_inversion}
\end{center}
\end{figure}

In all the experiments, the onset of the runaway greenhouse is related to the appearance of temperature inversion on the day side, away from the terminators (Fig.~\ref{fig_inversion}), followed by the collapse of convection and clouds, similar to that found in \cite{Wolf_2015}, \cite{Popp_2015,Popp_2016}, and \cite{Kopparapu_2017}. As shown in Fig.~\ref{fig_inversion}(a, c, \& e), in the experiment of 0.25 bar and 1750~W\,m$^{-2}$, temperature inversion and  cloud collapse occur in the 6th model year within the region of 30$^\circ$--60$^\circ$ (in tidally locked coordinates); after that, the planetary albedo decreases dramatically. So, the global-mean shortwave absorption becomes larger than the outgoing longwave radiation, and the system suddenly enters a runaway greenhouse state. Similar phenomena occur in the experiment of 1.0 bar and 1950~W\,m$^{-2}$ (Fig.~\ref{fig_inversion}(b, d, \& f)).

Sensitivity tests show that the onsets of the temperature inversion and cloud collapse do not depend on the initial condition (Fig.~\ref{fig_inversion}(g--j)). When the model is initialized from different surface temperatures, all the experiments enter runaway greenhouse and the curves of planetary albedo (as well as energy imbalance and outgoing longwave radiation, not shown) as a function of global-mean surface temperature roughly follow the same line (Fig.~\ref{fig_inversion}(i--j)).

The onset of the temperature inversion is due to strong shortwave absorption by water vapor when the low-level atmosphere has already become optical thick in thermal radiation \citep{Wordsworth_2013,Wolf_2015}. At high temperatures, water vapor strongly absorbs stellar radiation especially in the near-infrared wavelengths. For $p$N$_2$\,=\,0.25~bar, the system enters runaway greenhouse at a lower stellar flux threshold than that of 1.0 bar $p$N$_2$. This is mainly due to the effects of lapse rate and heat capacity as described above. More water vapor can be hold in the atmosphere and the heat capacity is smaller, causing a larger shortwave heating rate, so the onset of the temperature inversion occurs at a lower surface temperature ($\approx$280 versus 320~K in global mean, see Fig.~\ref{fig_inversion}(i--j)). For $p$N$_2$\,=\,4.0~bar, the system also enters runaway greenhouse at a lower stellar flux than that of 1.0 bar. This is due to the increases of relative humidity (as discussed in Section~\ref{sec_climate}) and of water vapor concentration;  the latter is associated with the combined effect of pressure broadening and water vapor feedback. As $p$N$_2$ is increased, the greenhouse effect becomes stronger due to pressure broadening, so surface and air temperatures increase and then more water vapor can be hold in the atmosphere. Because water vapour is a greenhouse gas, the greenhouse effect further raises, which leads to even greater surface and air warming and larger shortwave heating rate, promoting the onsets of the temperature inversion and cloud collapse.


\section{Summary}\label{summary}

\begin{figure}
    \includegraphics[width=\linewidth]{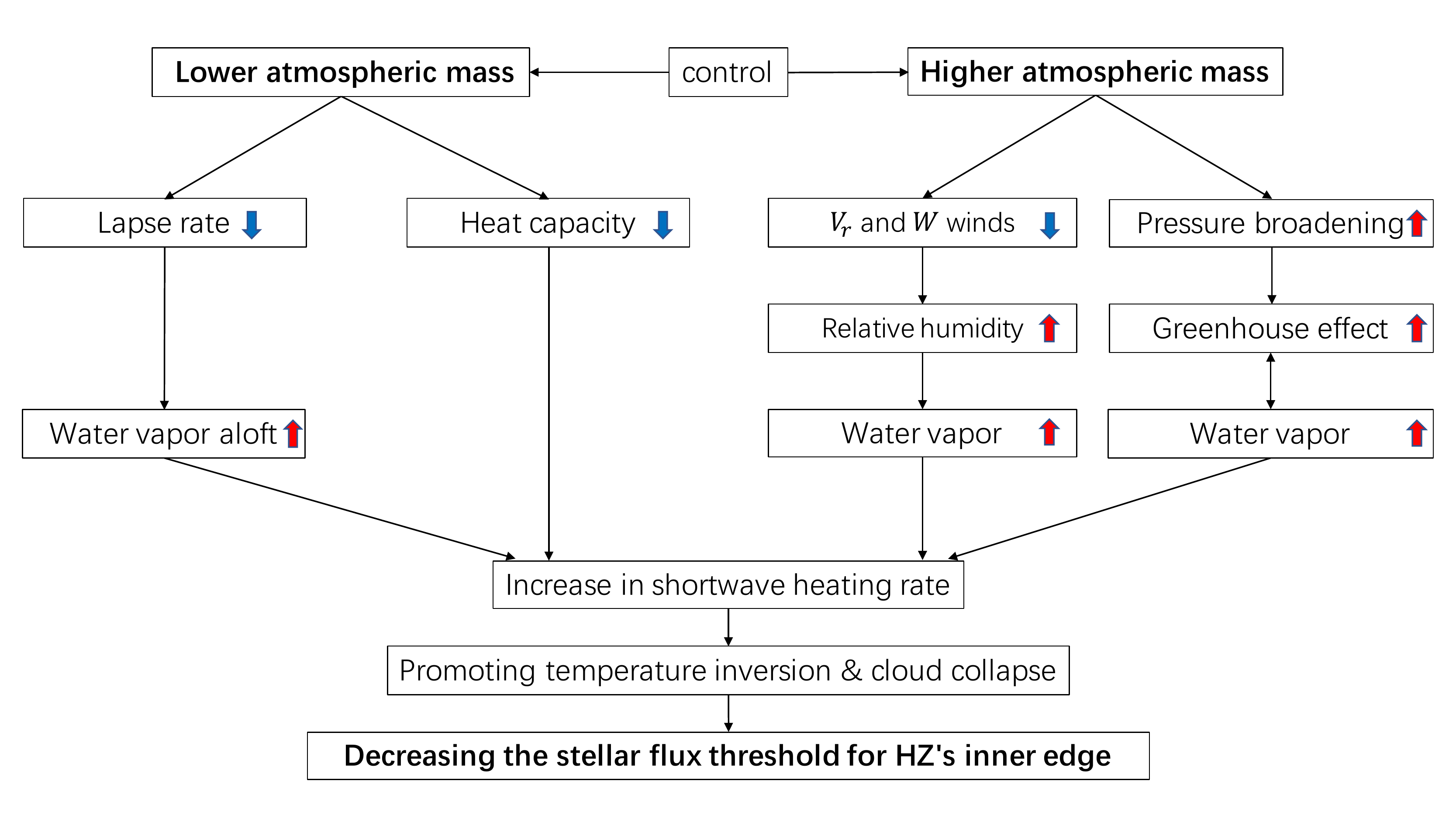}
    \caption{Schematic illustration for the effects of varying N$_2$ partial pressure on the inner edge of the habitable zone. The inner edge is defined based on runaway greenhouse. Five factors, pressure broadening, heat capacity, lapse rate, relative humidity, and clouds, make the problem be much more complex than that found in 1D models and cause the stellar flux threshold for the runaway greenhouse onset to be a non-monotonic function of $p$N$_2$.}
    \label{fig_initial}
\end{figure}

The 3D global climate model ExoCAM was employed to investigate the effects of varying $p$N$_2$ on the stellar flux threshold for the onset of the runaway greenhouse state on tidally locked rocky planets around M dwarfs. Comparing previous studies on this problem using 1D radiative-convective models \citep{Nakajima_1992,Kasting_1993,Goldblatt_2009, Kopparapu_2014,Ramirez_2020}, main new findings are summarized as follows and schematically shown in Fig.~\ref{fig_initial}.

   \begin{enumerate}
    \renewcommand{\labelenumi}{(\theenumi)}
      \item The global-mean surface temperature and the stellar flux threshold for the runaway greenhouse onset are non-monotonous functions of $p$N$_2$, due to the competing effects of five different processes, including clouds, pressure broadening, heat capacity, lapse rate, and relative humidity. Rayleigh scattering is unimportant for the inner edge, due to that cloud albdeo and near-infrared absorption by water vapor are effective in masking the effect of Rayleigh scattering.

      \item Lapse rate and water vapor aloft decrease with increasing $p$N$_2$ and atmospheric heat capacity increases with $p$N$_2$, so shortwave heating rate by water vapor decreases with $p$N$_2$ under a given surface temperature. These act to delay the onsets of temperature inversion and runaway greenhouse.

      \item The effects of pressure broadening and N$_2$--N$_2$ absorption increase with $p$N$_2$; this warms the surface and increases water vapor concentration. Water vapor feedback further amplifies the warming. These promote the onsets of temperature inversion and runaway greenhouse.
      
      \item The atmospheric circulation (W and V$_r$ winds) is a clear monotonically decreasing function of $p$N$_2$ although the horizontal energy transport increases with $p$N$_2$. This can influence the relative humidity. But, no clear trend is found in cloud fraction or cloud albedo on the day side, due to complex moist processes and the interactions between them and atmospheric circulation, although the night-side clouds exhibit a clear decreasing trend with $p$N$_2$. 

      \item Finally, for the inner edge, the magnitude for the effect of varying $p$N$_2$ is within 13\% under the parameters we examined, which is comparable to that of varying planetary radius and gravity (within $\approx$9\%, \cite{Yang_2019_effects}) and of the uncertainty in pure water vapor radiation transfer (within $\approx$10\%, \cite{Yang_2016}), but smaller than that of the uncertainty in cloud scheme (within $\approx$50\% in \cite{Bin_2018}) and of varying rotation period and stellar spectrum (within $\approx$70\%, \cite{Kopparapu_2017}).
   \end{enumerate}

Future work is required to investigate tidally locked planets but in spin-orbit resonance states like Mercury and rapidly rotating planets like Earth. For these planets, atmospheric circulation and cloud distribution are different from those of 1:1 tidally locked planets \citep{Yang_2014,Wang_2016,Salameh_2018}; this can strongly influence the trend of the effect of $p$N$_2$ on the inner edge, following the analyses above. Finally, we note that atmospheric masses or N$_2$ pressures on exoplanets may could be inferred from the observations of phase curves \citep{Koll_2015,Kreidberg_2019,Koll_2019}, emission and  transmission spectra (especially the N$_2$-N$_2$ absorption in 4.15 $\mu$m and in the wings of the 4.3 $\mu$m CO$_2$ band; \cite{Schwieterman_2015}), or Raman scattering \citep{Oklopvcic_2016}.


\acknowledgments

We are grateful to Eric Wolf for the release of the model ExoCAM and to Ravi Kumar Kopparapu for his helpful comments and suggestions. J.Y. acknowledges support from the National Natural Science Foundation of China (NSFC) under grant 41675071.

\software{ExoRT: https://github.com/storyofthewolf/ExoRT}
\software{ExoCAM: https://github.com/storyofthewolf/ExoCAM}
\software{Tidally Locked Coordinates: https://github.com/ddbkoll/tidally-locked-coordinates}


\bibliography{misc}{}
\bibliographystyle{aasjournal}


\newpage

\end{document}